\documentclass{JINST}

\usepackage{amssymb}

\usepackage{graphics}
\usepackage{graphicx}

\title{Detector for imaging of explosions: present status and future
prospects with higher energy X-rays.}

\author{Aulchenko V.M.$^a$,Evdokov O.V.$^b$, Shekhtman L.I.$^a$\thanks{Corresponding author},Ten K.A.$^c$,
Tolochko B.P.$^b$, Zhogin I.L.$^b$, Zhulanov V.V.$^a$ \\
\llap{$^a$}Budker Institute of
Nuclear Physics\\ 11 Lavrentiev Avenue, Novosibirsk 630090\\
Russia. Fax: 7(383)3307163,\\ e-mail: L.I.Shekhtman@inp.nsk.su\\
\llap{$^b$}Institute of Solid-State Chemistry and Mechano-Chemistry\\
630090 Novosibirsk, Russian Federation\\
\llap{$^c$}Lavrentiev Institute of Hydrodynamics\\
630090 Novosibirsk, Russian Federation}

\abstract {The detector for imaging of explosions (DIMEX) is in
operation at the synchrotron radiation (SR) beam-line at VEPP-3
electron ring at Budker INP since 2002. DIMEX is based on
one-coordinate gas ionization chamber filled with Xe-CO$_2$(3:1)
mixture at 7atm, and active Frisch-grid made of Gas Electron
Multiplier (GEM). The detector has spatial resolution of
$\sim$0.2mm and dynamic range of $\sim$100 that allows to realize
the precision of signal measurement at a percent level. The frame
rate can be tuned up to 8 MHz (125 ns per image) and up to 32
images can be stored in one shot. At present DIMEX is used with
the X-ray beam from 2T wiggler that has $\sim$20 keV average
energy. Future possibility to install similar detector at the SR
beam-line at VEPP-4 electron ring is discussed.  }


\begin{document}

\section{Introduction.}

The detector for imaging of explosions (DIMEX) is in operation at
the synchrotron radiation (SR) beam-line at VEPP-3 electron ring
at Budker INP since 2002 (~\cite{DIMEX1}, ~\cite{DIMEX2},
~\cite{DIMEX3}). DIMEX is used for the studies of material
properties in the conditions with very high temperatures and
pressures (i.e. during detonation) as well as of the processes of
formation of new materials (including nano-structures) during an
explosion (~\cite{appDIMEX1}, ~\cite{appDIMEX2}). The studies are
performed either by the measurements of absorption of the SR beam
by an exploding sample (direct absorption experiments), or by the
measurements of photon flux scattered at small angles from an
exploding sample (SAXS experiments). In both types of experiments
DIMEX allows the measurement of one-coordinate distribution of the
X-ray flux emitted by a single electron bunch and then either
partially absorbed or scattered by a sample. The detector measures
a sequence of up to 32 of such images corresponding to subsequent
bunches. Thus the effective time resolution of the method is
determined by the length of an electron bunch in the accelerator
and is below 1ns.

At present DIMEX is operating at the beam-line with white SR beam
from 2T wiggler with average energy of photons around 20keV (after
passing through Be windows (5mm) and the sample (1cm of
explosive))~\cite{DIMEX1}. This low energy imposes limit on the
sample thickness of $\sim$1cm because otherwise the absorption in
the sample becomes two strong and limited statistics of
transmitted photons does not allow to get necessary precision of
the measurement. Higher energy of SR beam can be obtained at
VEPP-4M with 5-pole 1.3T wiggler that is now under development.
The dedicated simulation study has been performed to get estimates
of the main parameters of DIMEX for the X-ray spectrum at VEPP-4M.
The present paper describes current status of DIMEX performance at
VEPP-3 and summarizes the results of simulation studies of
possible operation at VEPP-4M.

\section{Present status of DIMEX at VEPP-3. }

\begin{figure}[htb]
\centering
\includegraphics[width=0.7\textwidth,clip]{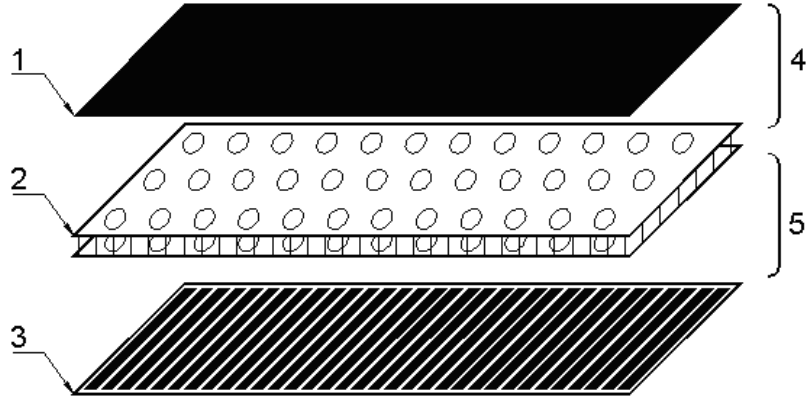}
\caption{Schematic view of DIMEX design . 1 - drift electrode, 2 -
GEM, 3 - strip board, 4 - drift gap, 5 - induction gap.}
\label{fig:DIMEXdesign}
\end{figure}

DIMEX is based on the high pressure gas ionization chamber with
active Frisch-grid made of gas electron multiplier (GEM). Design
of the detector is shown schematically in
Fig.~\ref{fig:DIMEXdesign}. X-ray beam enters the detector box
through carbon fiber window and is absorbed in the gap between the
drift electrode and top GEM side. Electrons of primary ionization
drift towards GEM, partially penetrate through it and then drift
through the gap between bottom GEM side and strip board (induction
gap). During the last phase the electrons induce charge at the
strips (strip pitch is 0.1mm). Detector is filled with Xe-CO$_2$
(3:1) mixture at 7 atm (absolute).

Each strip of the anode board is connected to the input of the
integrator chip APC128~\cite{APC128}. This chip contains 128
channels with integrator and 32-cell analogue pipeline in each
channel. The charge from the integrator can be stored in any of
the pipeline cells. The pipeline can be read out through the
analogue multiplexer. The pipeline switching and readout frequency
can be up to 10MHz.

GEM shields the strip board from the charge induction from
positive ions and absorbs part of the electrons passing through
thus reducing the signal from each photon. As the dynamic range of
the APC128 integrator is limited, the reduction of charge induced
per single photon increases effectively the dynamic range in terms
of photons. At present the GEM transparency is about 0.5 with the
drift field of 16kV/cm, induction field of 8kV/cm and GEM voltage
of 800V (for example, see data from ~\cite{GEMtransp1} and
~\cite{GEMtransp2}). The transparency can be changed within a
certain limits and thus the dynamic range can be tuned. However we
have not yet tried this option.

Detailed description of the operation algorithm of DIMEX can be
found in ~\cite{DIMEX3}. The detector is opened to the beam for
$\sim$60$\mu$s by a special rotation shutter and explosion is
triggered during this time gap. The exact synchronization between
the detonation and the detector timing sequence is achieved with
the help of two wires embedded into the explosive. The wires are
shorted when the detonation wave is passing through them.

Bunch crossing time of VEPP-3 synchrotron is 250ns in 1-bunch
regime. The accelerator can also operate in 2-bunch mode. The
timing sequence of DIMEX is synchronized with the bunch crossing.
Data from up to 32 bunches can be stored separately in the
pipeline cells. After recording the data from 32 bunches the
detector sequence is stopped and data from the pipeline is
transferred to the computer through 100Mb/s ethernet line.

\begin{figure}[htb]
\centering
\includegraphics[width=0.8\textwidth,clip]{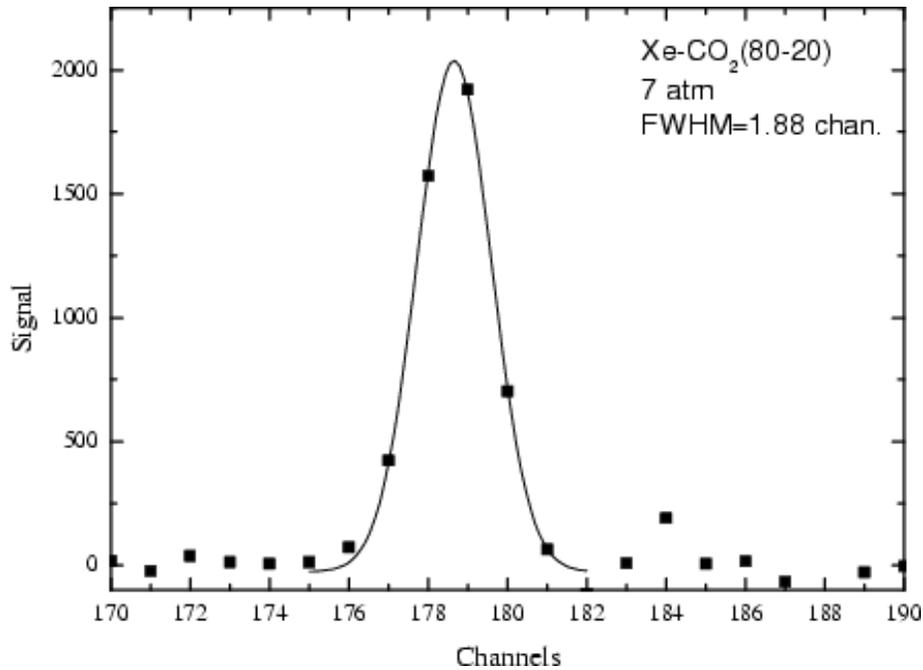}
\caption{The image of thin beam (line spread function) measured
with the edge method. }
\label{fig:Spatial_res}
\end{figure}

Spatial resolution of DIMEX is determined mostly by pressure of
the operating gas mixture. The line spread function measured by
the edge method is shown in Fig.~\ref{fig:Spatial_res}. The
resolution (FWHM) is close to 0.2mm.

\begin{figure}[htb]
\centering
\includegraphics[width=0.8\textwidth,clip]{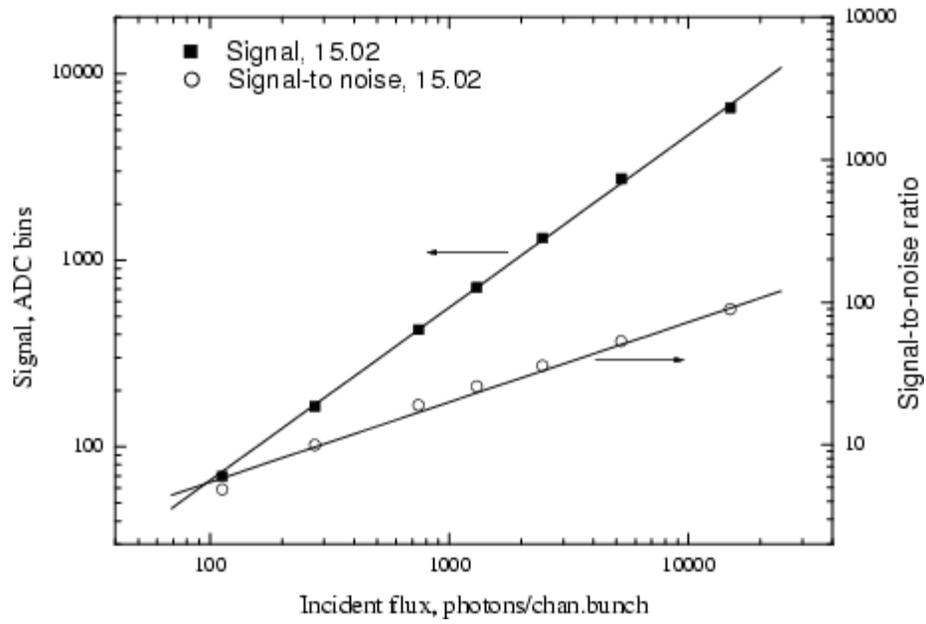}
\caption{Signal (left scale) and signal to noise ratio (right
scale) as a function of the X-ray flux at the entrance window of
the detector.}
\label{fig:Signal}
\end{figure}

Signal and signal to noise ratio as a function of incident flux,
calculated from the known electron beam energy, current and
B-field of the wiggler\footnote{Calculated with XOP2.0} are shown
in Fig.~\ref{fig:Signal}. Signal to noise ratio reaches $\sim$100
demonstrating the possibility of signal measurement at a percent
level of precision. The maximal signal is limited by the space
charge accumulation due to slow positive ions in the drift gap.

\begin{figure}[htb]
\centering
\includegraphics[width=0.8\textwidth,viewport=10 10 500 450,clip]{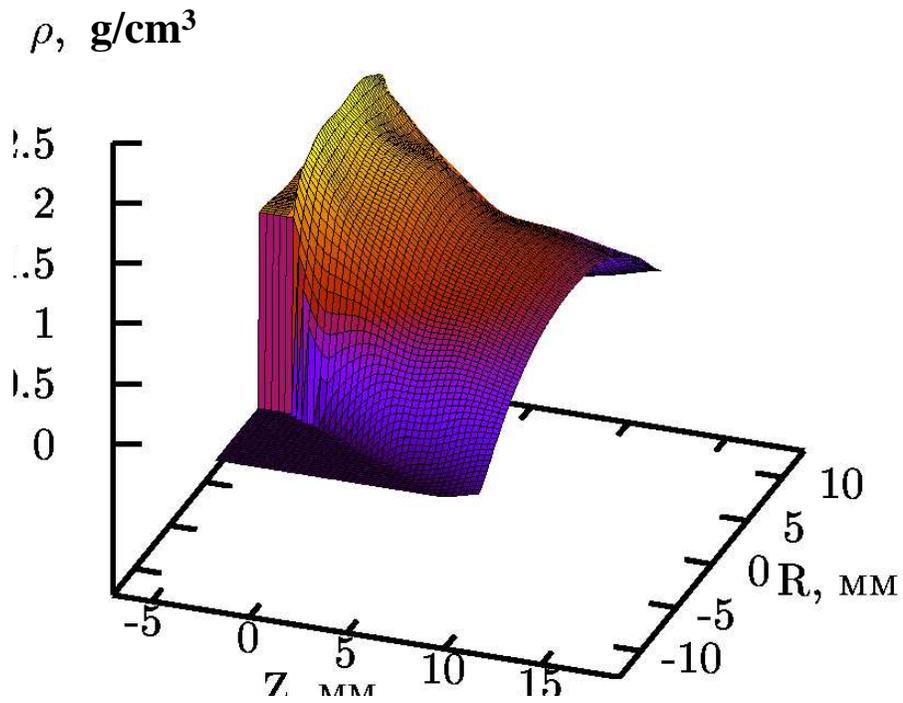}
\caption{Reconstructed density map of the exploding sample.} \label{fig:Density}
\end{figure}

The result of one of the direct absorption experiments is shown in
Fig.~\ref{fig:Density}, where the density map of the exploding
sample is reconstructed ~\cite{appDIMEX1}. The data are
reconstructed from the time evolution of the transmitted signal in
one slice, assuming constant speed of the detonation wave. In the
figure we can see the undistorted sample at Z<-5mm with the
density of $\sim$1.5g/cm$^3$. At Z=-5mm the density jumps above
$\sim$2.0g/cm$^3$ due to the detonation and then the sample is
decaying with the density steadily decreasing. The most important
observation in this kind of experiments is the value of density
increase behind the detonation wave, and the exact density profile
at the moment of detonation.

\begin{figure}[htb]
\centering
\includegraphics[width=0.8\textwidth,clip]{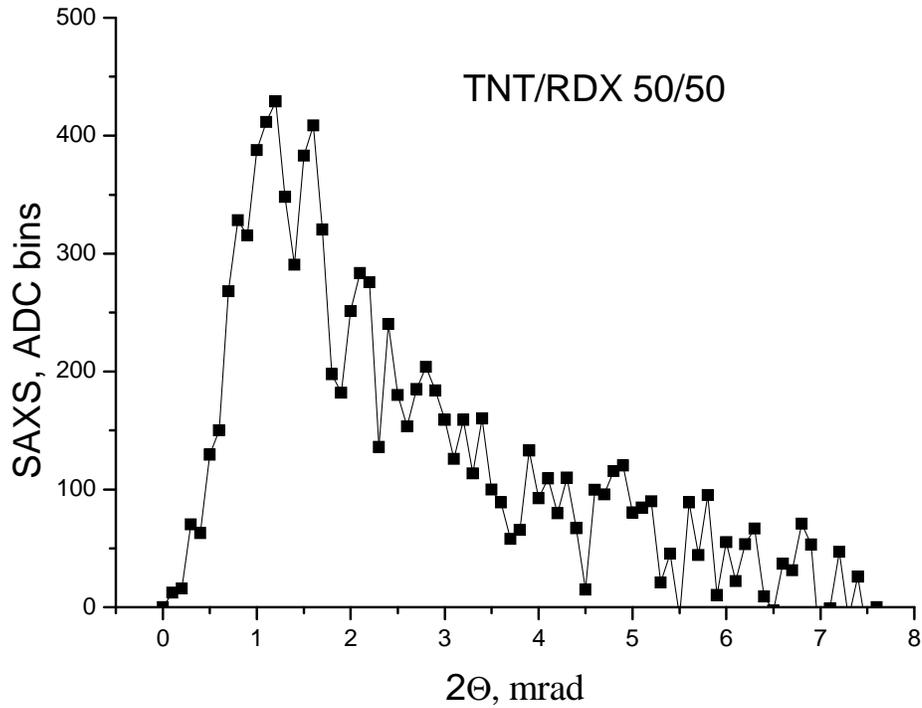}
\caption{SAXS experiment with the explosion of composite sample.
Angular distribution of SAXS radiation 3$\mu$s after the passage
of detonation wave.} \label{fig:SAXS1D}
\end{figure}

\begin{figure}[htb]
\centering
\includegraphics[width=0.8\textwidth,clip]{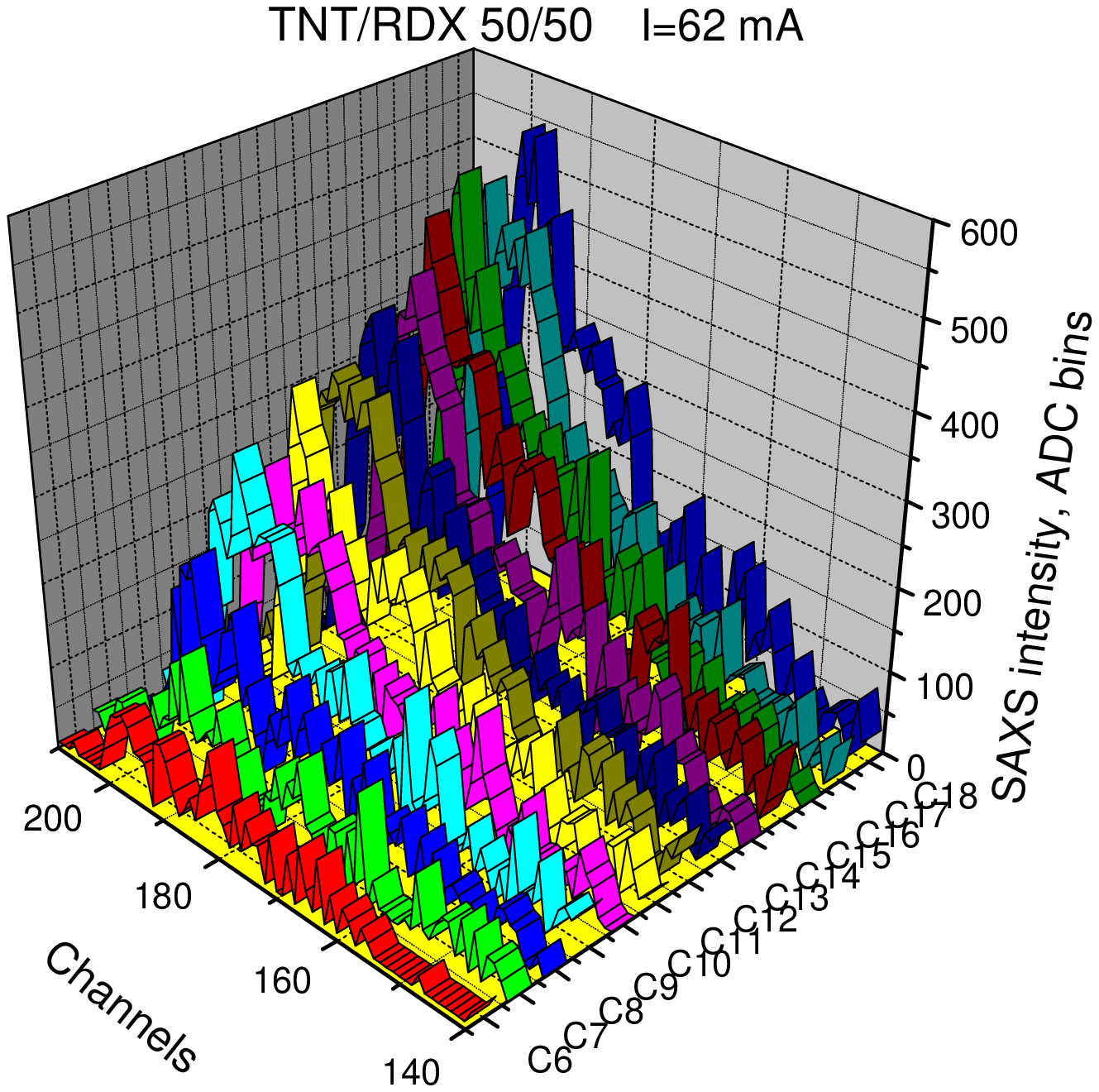}
\caption{Time evolution of the angular distribution of SAXS.
Images from different frames have different colors. Frame rate is
2MHz. } \label{fig:SAXS2D}
\end{figure}

SAXS experiments can give information about the amount and size of
particles that are produced in an explosion. The example of SAXS
experiment is shown in Fig.~\ref{fig:SAXS1D} and
Fig.~\ref{fig:SAXS2D} ~\cite{appDIMEX2}. In Fig.~\ref{fig:SAXS1D}
the angular distribution of scattered X-rays is shown 3$\mu$s
after the detonation wave passed through the slice of the sample
exposed to the beam. In Fig.~\ref{fig:SAXS2D} the whole evolution
of SAXS angular distribution in time is shown. We can see that
SAXS intensity is growing with time for quite a long period after
the detonation.

After the first years of DIMEX operation the power of the method
of imaging of the radiation from single electron bunch has become
obvious. However further improvement of the detector is necessary.
For the direct absorption experiments the improvement of spatial
resolution is needed down to 50$\mu$m and below as well as the
increase of dynamic range and the precision of signal measurement.
For both SAXS and direct absorption experiments the higher frame
rate would be useful that is however limited by the bunch timing
of the accelerator. Also for both SAXS and direct absorption
experiments the transition to higher X-ray energies would be
useful. Higher X-ray energies will allow to study thicker samples
with higher densities.

\section{Prospects of DIMEX with higher energy SR at VEPP-4M.}

\begin{figure}[htb]
\centering
\includegraphics[width=0.7\textwidth,clip]{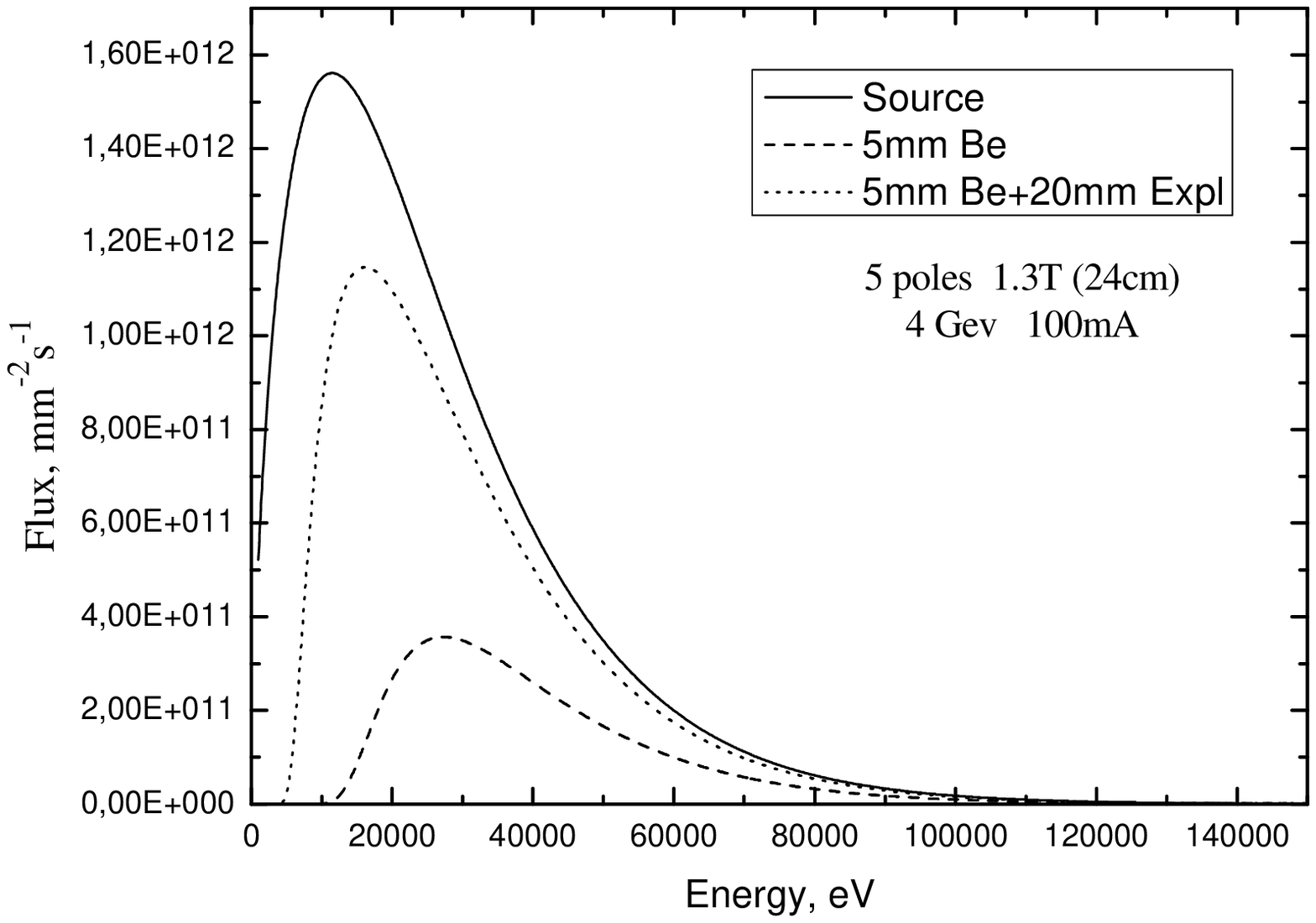}
\caption{Spectrum of radiation from 5-pole 1.3T wiggler at
VEPP-4M.} \label{fig:Spectrum}
\end{figure}

The energy spectrum of radiation from the 5-pole 1.3T wiggler that
is now under development for VEPP-4M ring, is shown in
Fig.~\ref{fig:Spectrum}\footnote{Spectrum is calculated with
XOP2.0}. The figure presents the primary spectrum, the spectrum
after absorption in Be windows with total thickness of 5mm and the
spectrum after absorption in 20mm of explosive (dashed line, below
will be referred as spectrum 3, pure carbon at a density of
1.6g/cm$^3$ has been taken as an explosive). We can see that
spectrum 3 has maximum at $\sim$30 keV. At around 50 kev its
spectral density is about half of the maximum and the tail spreads
up to 100 keV.

\begin{figure}[htb]
\centering
\includegraphics[width=0.7\textwidth,viewport=10 10 500 450,clip]{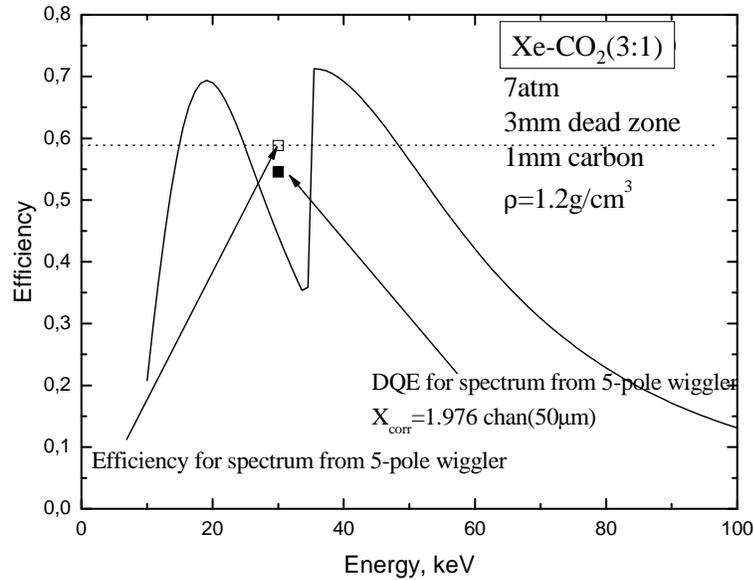}
\caption{Efficiency as a function of energy. Efficiency and DQE
for the radiation with spectrum 3 shown at E=30keV. } \label{fig:EffDIMEX}
\end{figure}

The efficiency of detector with 1mm carbon fiber window of 1.2
g/cm$^3$ density, dead zone of gas 3mm in depth and 30mm deep
sensitive zone, is shown in Fig.~\ref{fig:EffDIMEX} as a function
of X-ray energy. The efficiency has been calculated for the gas
mixture of Xe-CO$_2$(3:1) at 7atm absolute pressure (calculation
has been performed with XOP2.0). By efficiency we mean here the
probability of absorption of X-ray photon in the sensitive region
of the detector.

The efficiency of DIMEX to photons with spectrum 3 has been
calculated by Monte-Carlo simulation. Photon flux distributed
uniformly along the direction of entrance window was directed
perpendicular to the window plane. The photons energy was
distributed according to spectrum 3. Photons that entered the
detector volume have released their energy partly or totaly in the
dead zone or in the sensitive region. Part of the photon flux
transferred through the detector volume without interaction. The
simulation has been performed with FLUKA2006.3 package
~\cite{FLUKA1},~\cite{FLUKA2}. In order to determine the
probability of interaction in the sensitive region the number of
photons entering this region have been detected as well as the
number of primary photons leaving the sensitive region without
interaction. The difference between the former and the latter
normalized by the number of primary photons has given the
efficiency.

Apart from the probability of interaction of photons distributed
following spectrum 3, the simulation has given the value of
detective quantum efficiency(DQE) that characterizes the
possibility to detect small deviations of signal at noisy
background and with limited photons statistics
~\cite{DQE1},~\cite{DQE2}. DQE can be determined as a square of
signal to noise ratio normalized to the input flux (in photons per
channel) and to the correlation interval $X_{corr}$ (in channels),
where $X_{corr}$ is FWHM of the autocorrelation function.

\begin{equation}\label{DQE1}
    DQE = (s/n)^2/(N\centerdot~X_{corr})
\end{equation}

\begin{figure}[htb]
\centering
\includegraphics[width=0.7\textwidth,viewport=10 10 500 450,clip]{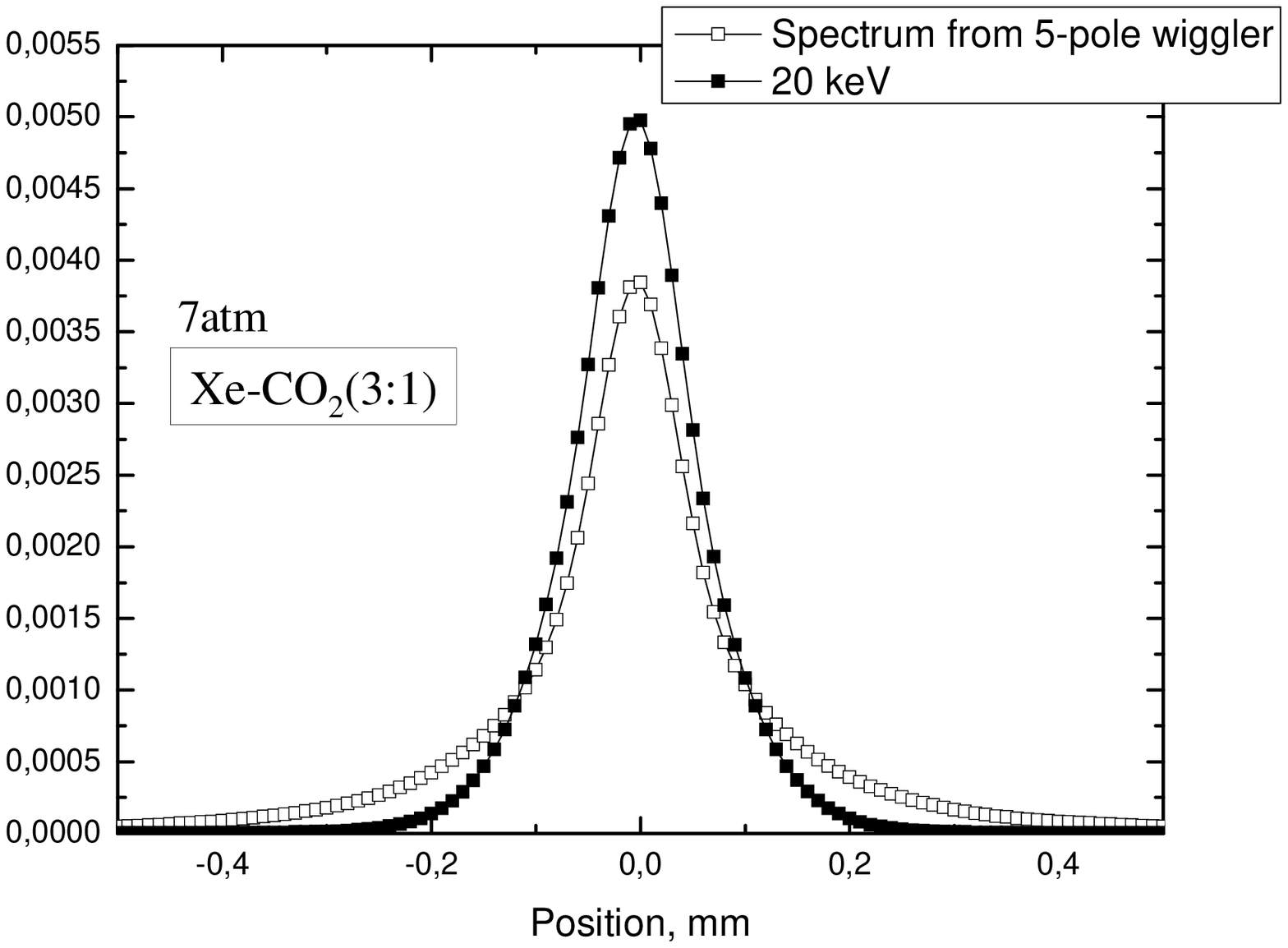}
\caption{Comparison of LSF for 20keV photons and photons with
energy spectrum 3 with diffusion.} \label{fig:LSF20kevSp3Diff}
\end{figure}

For the calculation of DQE the distribution of energy deposited in
the sensitive region along the direction of the input window has
been simulated. Uniform X-ray flux has been falling onto the
detector within 1cm region along the input window. The whole
region has been divided into 200 50$\mu$m intervals. The rms and
mean values of energy deposited in these intervals have been
calculated. The former has been taken as noise and the latter as
signal value. The input flux has been known from simulation
conditions. For the simulation of the autocorrelation function the
pencil-like beam has been used with all the photons entering
detector at a constant coordinate. As a result the line-spread
function (LSF) of the detector has been obtained (see
Fig.~\ref{fig:LSF20kevSp3Diff}). The autocorrelation function is a
square of LSF. As a result of simulation the efficiency obtained
is equal to 59$\%$ and DQE is equal to 48$\%$. DQE is less than
the probability of absorption in the sensitive region because of
fluctuations of energy deposited after the photons absorption.

\begin{figure}[htb]
\centering
\includegraphics[width=0.7\textwidth,viewport=10 10 500 450,clip]{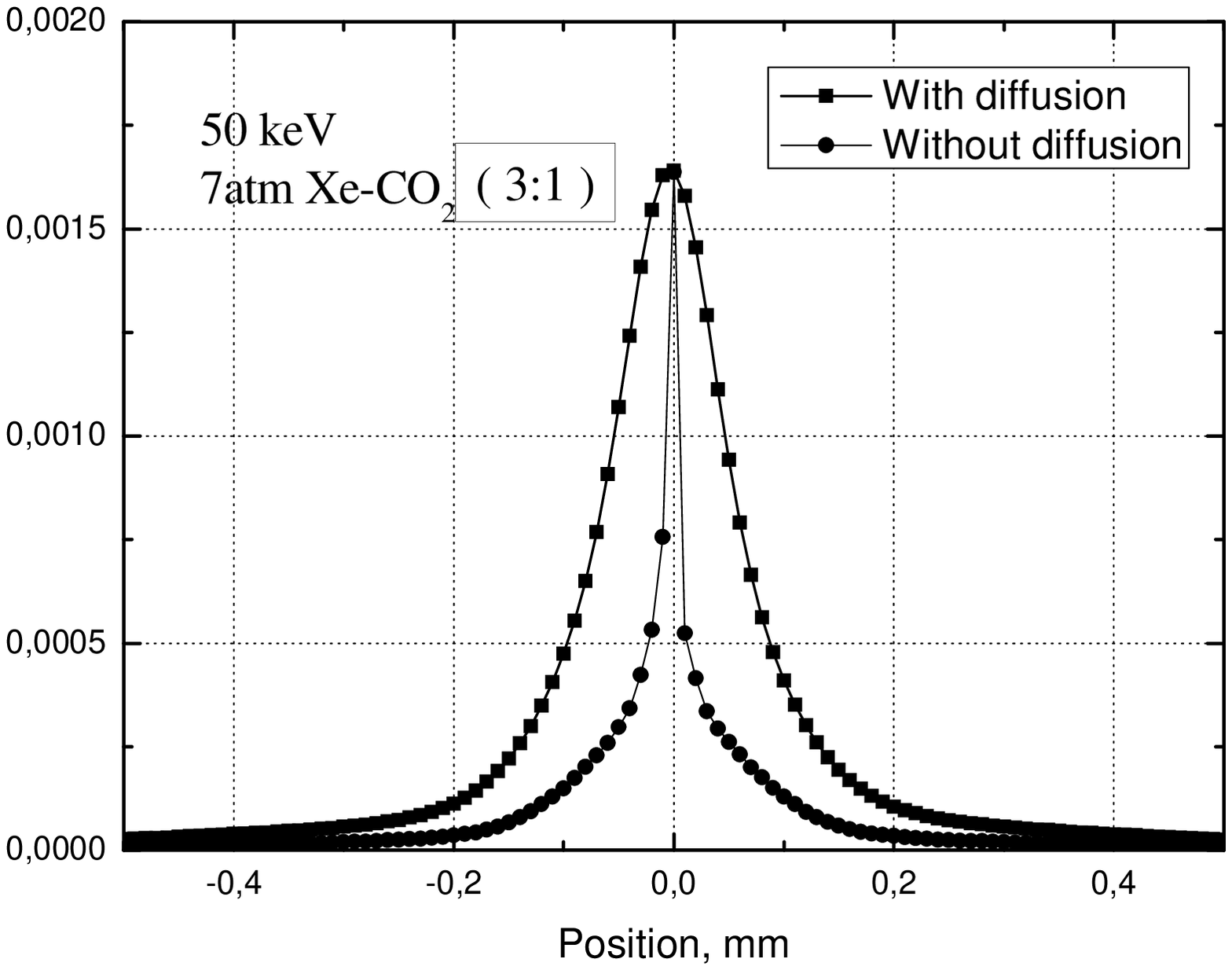}
\caption{Line spread function for 50keV X-rays. Comparison of the
cases with diffusion and without diffusion.} \label{fig:Diffusion}
\end{figure}

Spatial resolution of DIMEX has been characterized by LSF that was
obtained by the simulation. Different processes affecting the
resolution could be included. In Fig.~\ref{fig:Diffusion} the
comparison between LSF with and without electrons diffusion is
shown for the beam of 50keV photons. The drift length in the
detector is 2.5mm and the variance of electron position due to
diffusion is 33$\mu$m. Fig.~\ref{fig:Diffusion} demonstrates that
spatial resolution of DIMEX is mostly determined by diffusion.

\begin{figure}[htb]
\centering
\includegraphics[width=0.7\textwidth,viewport=10 10 500 450,clip]{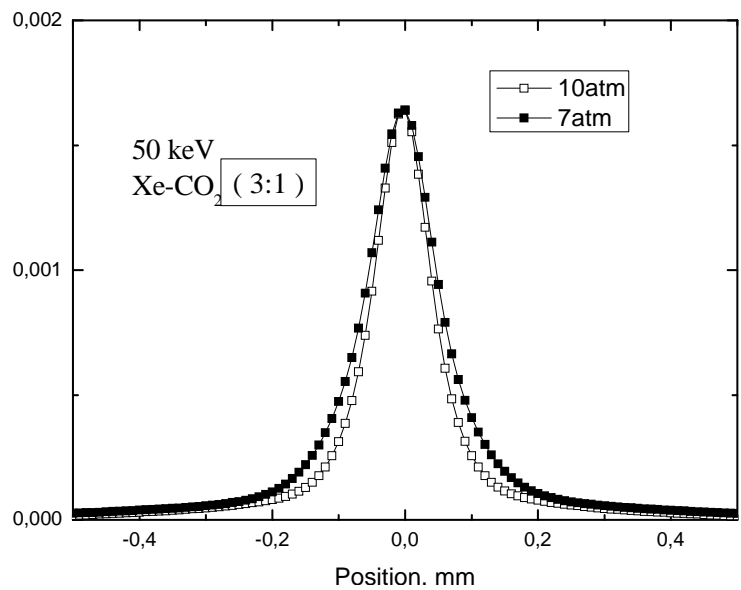}
\caption{Modification of LSF with pressure.} \label{fig:Pressure}
\end{figure}

The influence of pressure on the spatial resolution of DIMEX is
shown in Fig.~\ref{fig:Pressure}. The improvement of spatial
resolution with pressure is determined by the diffusion that is
reversed proportional to square root of pressure. Thus the change
of resolution is negligible when pressure is increased from 7atm
to 10atm. Further increase of pressure is problematic because of
too high voltages that has to be kept on the detector elements (in
order to keep constant E/p).

\begin{figure}[htb]
\centering
\includegraphics[width=0.7\textwidth,viewport=10 10 500 450,clip]{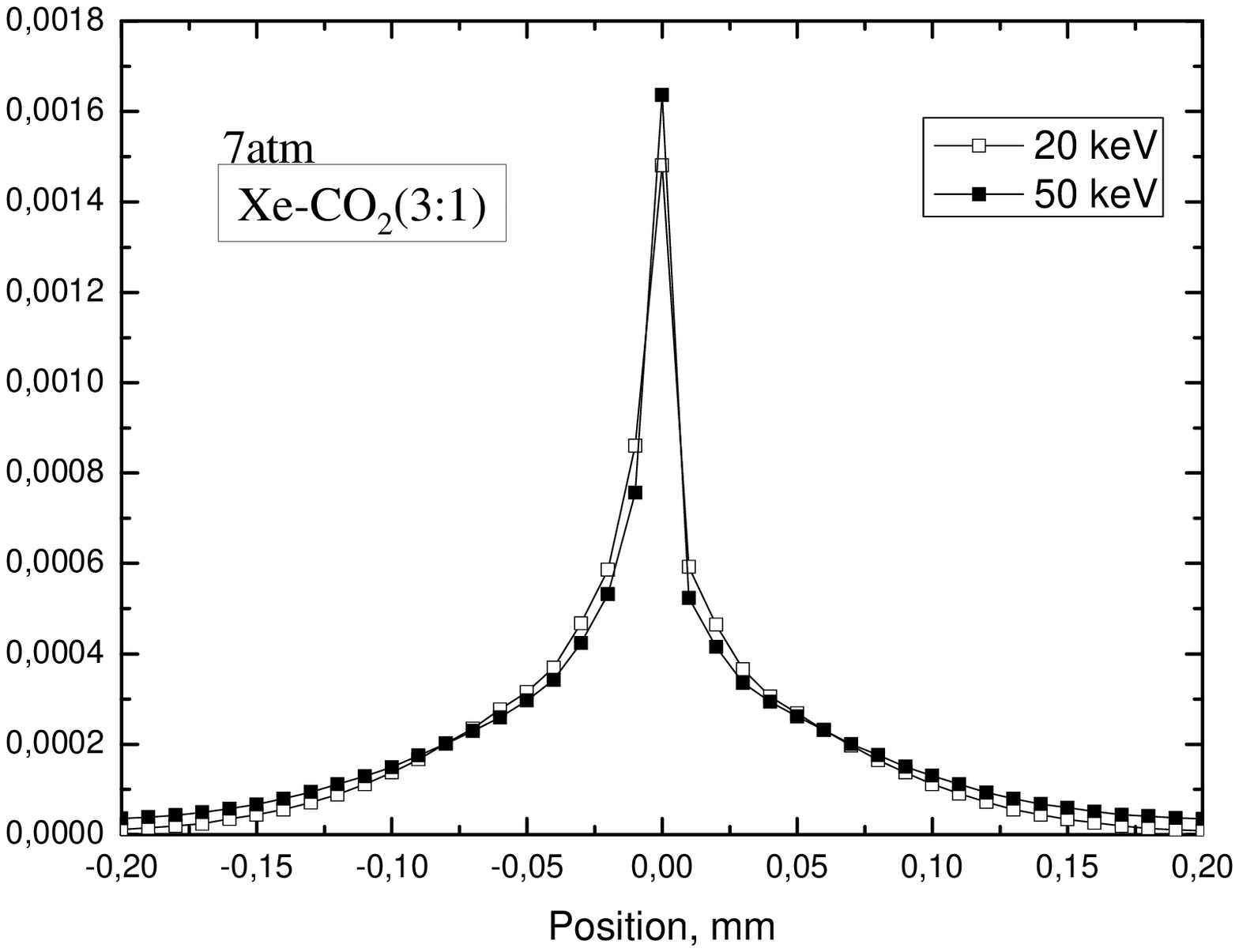}
\caption{Comparison of LSF for 20keV and 50keV photons. LSF is
simulated without diffusion.} \label{fig:EnergyLSF}
\end{figure}

\begin{figure}[htb]
\centering
\includegraphics[width=0.7\textwidth,viewport=10 10 500 450,clip]{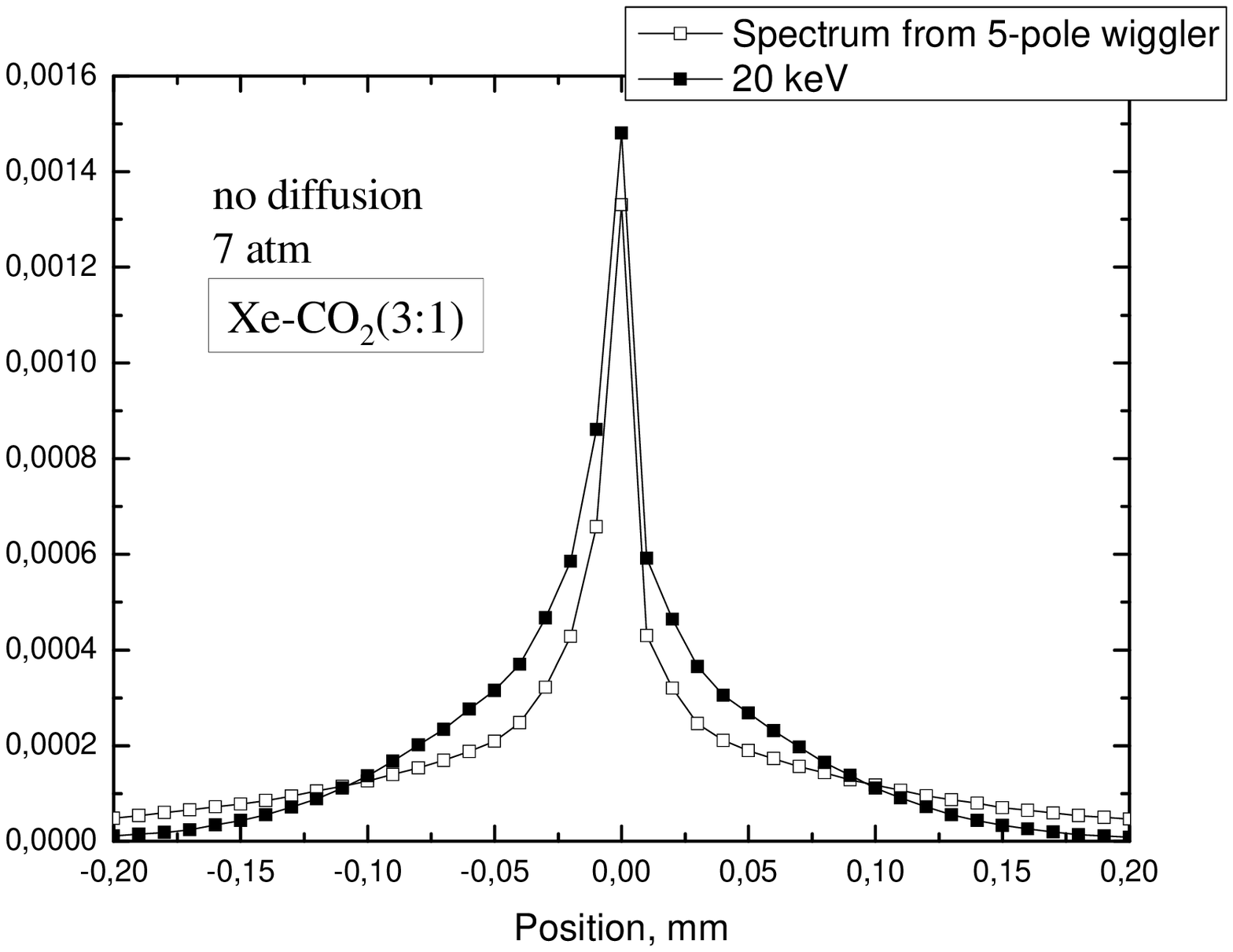}
\caption{Comparison of LSF for 20keV photons and photons with
energy spectrum 3 without diffusion.} \label{fig:LSF20keVSp3}
\end{figure}

Comparison of the spatial resolution at two different energies
20keV and 50keV (Fig.~\ref{fig:EnergyLSF}) demonstrates that the
dependence of LSF on energy is negligible in this energy range.
Spatial resolution for X-rays with energy distribution as in
spectrum 3 is shown in Fig.~\ref{fig:LSF20keVSp3} in comparison
with the spatial resolution for 20 keV photons. The line spread
functions in Fig.~\ref{fig:EnergyLSF} and ~\ref{fig:LSF20keVSp3}
are taken without diffusion to observe better the effect of
different photon energies. From Fig.~\ref{fig:LSF20keVSp3} we can
see that the LSF for 20 keV photons is wider in the central part
than the LSF for spectrum 3. It happens due to low energy
photoelectrons that are emitted after photo-effect  on K-edge of
Xe ($\sim$35keV). The LSF for spectrum 3 has much wider halo
around the central part with long tails as compared to the LSF for
20 keV photons that is formed by long high energy tail of the
energy spectrum and by the fluorescent photons that are emitted
after the absorption on Xe K-edge.

Complete line spread functions with the diffusion taken into
account are compared in Fig.~\ref{fig:LSF20kevSp3Diff} for 20keV
photons beam and for X-rays with energy spectrum 3. The difference
between these two LSF is only in wider halo for the case of
spectrum 3. FWHM in both cases is about 170$\mu$m that fits very
well with the experimental result (see
Fig.~\ref{fig:Spatial_res}).

\begin{figure}[htb]
\centering
\includegraphics[width=0.7\textwidth,viewport=10 10 500 450,clip]{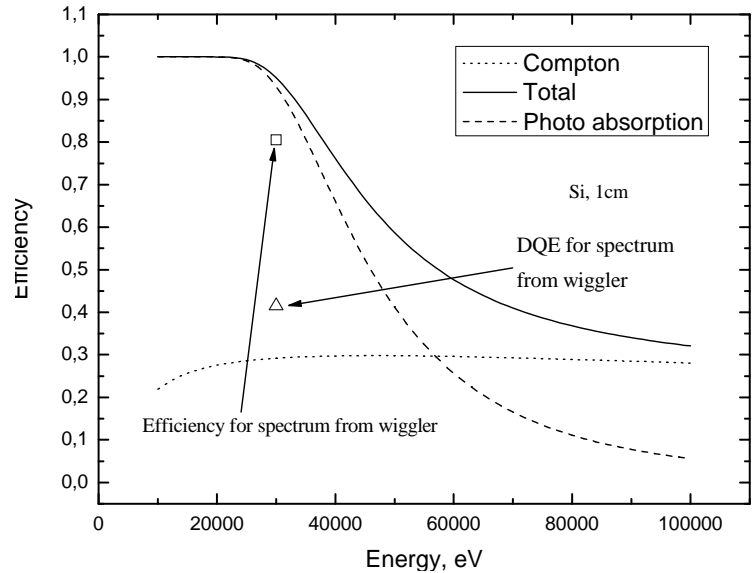}
\caption{Efficiency of 1cm thick Si detector as a function of
X-ray energy. Probability of a single interaction of one type
(Compton or photo-) and of any type. Efficiency for the X-ray beam
from 5-pole wiggler (spectrum 3) and DQE for the same radiation. }
\label{fig:EffSi}
\end{figure}

An alternative to the gaseous detector can be a solid-state
device. Fig.~\ref{fig:EffSi} shows the probability of interaction
through Compton scattering (at least one time), photo-absorption
and total probability of interaction in 1cm of silicon as a
function of X-ray energy. Below 30keV the probability of photon
absorption is close to 1. Compton scattering occurs with
probability of $\sim$30$\%$ in the energy range from 20keV to
100keV. The efficiency (probability of absorption or at least one
scattering) for radiation with spectrum 3 is also shown in the
figure as well as DQE for the same kind of X-rays.

As Compton scattering causes energy deposition through a recoil
electron, this process has been accounted for in the efficiency
calculation. The energy deposited in this way is much smaller than
that of primary photon and thus the deposited energy is
fluctuating significantly, that results in low DQE value. The
efficiency for photons with spectrum 3 in 1cm thick Si detector is
equal to $\sim$81$\%$ while DQE is $\sim$42$\%$.

\begin{figure}[htb]
\centering
\includegraphics[width=0.8\textwidth,viewport=10 10 500 450,clip]{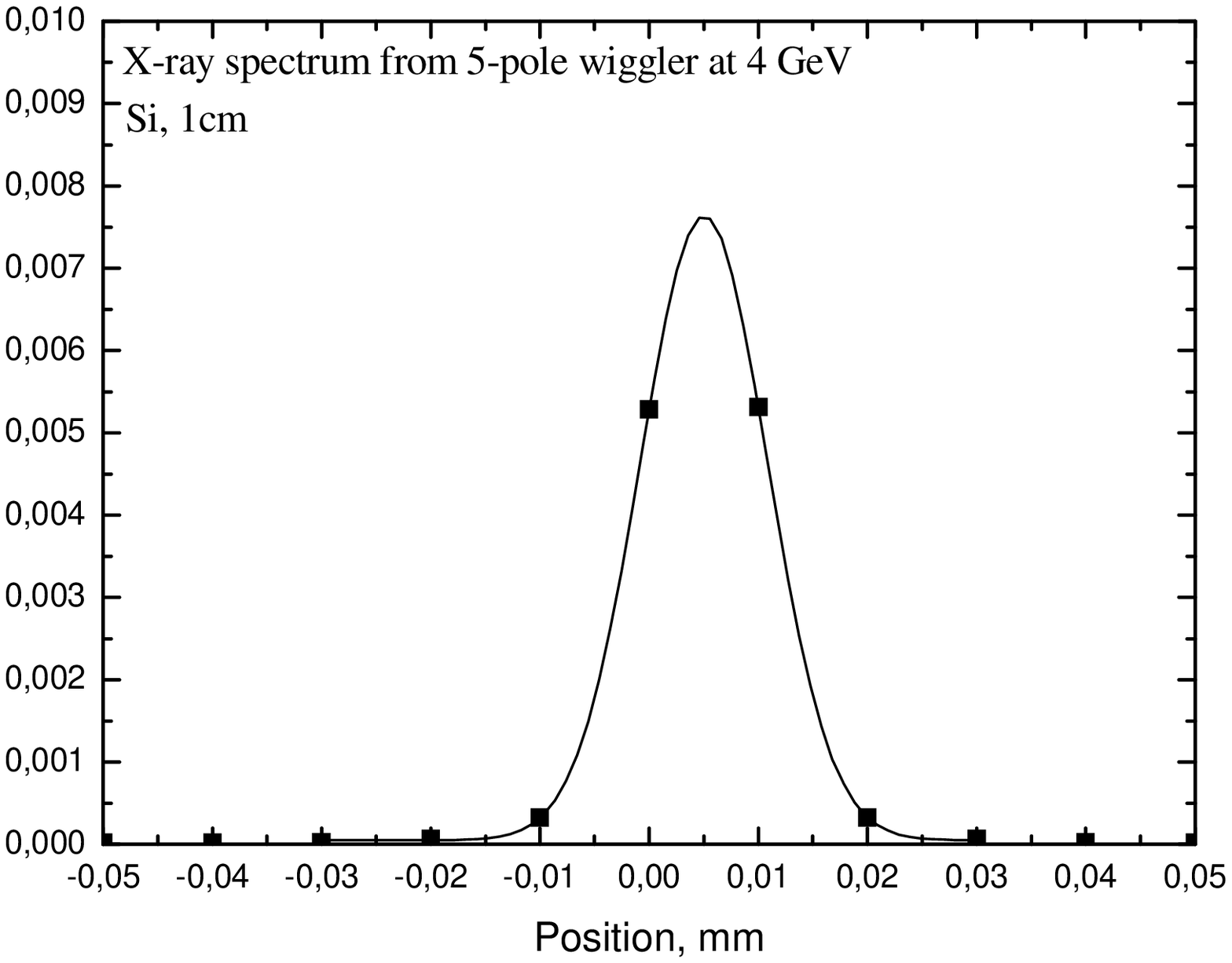}
\caption{Line spread function of 1cm thick Si detector for photons
with spectrum 3. } \label{fig:SiLSF}
\end{figure}

Line spread function of the Si detector has been obtained by
simulation with X-rays having energy distribution following
spectrum 3 (Fig.~\ref{fig:SiLSF}). Spatial resolution of this
detector is much better than of gaseous DIMEX. FWHM of the line
spread function is $\sim$20$\mu$m.

\section{Conclusions.}

5 years of operation of DIMEX at the SR beam-line at VEPP-3 has
proven that the method of imaging of radiation from separate
electron bunches is a very powerful tool. However in spite of
successful application of the detector to the direct absorption
imaging of detonation process and SAXS experiments, further
improvement of the method is desirable. The development of new
5-pole 1.3T wiggler at VEPP-4M and construction of the new
dedicated SR beam-line will allow the increase of energy and thus
the thickness of samples under study. The simulation study
performed in the present work demonstrates that the present
detector will not change its parameters with the higher energy SR
beam at VEPP-4M. The spatial resolution will stay  at $\sim$0.2mm
(FWHM) and DQE will be around 50$\%$.

Spatial resolution of DIMEX and maximal X-ray flux that this
detector can withstand are limited by the gaseous technology used
in the present device. Further improvement of both parameters can
be achieved by the development of solid-state DIMEX. Silicon
microstrip detector could be a good candidate if it is positioned
at a small angle to the beam. The simulation shows that if the
beam crosses Si detector within 1cm length, the DQE for X-ray
spectrum from VEPP-4M wiggler is close to that of the gaseous
DIMEX. The spatial resolution is however can be much better if
proper segmentation will be done, because the line spread function
of the Si detector before the application of any strip readout
structure is about 20$\mu$m wide (FWHM).

\end{document}